\documentstyle[12pt,aasms4]{article}

\def\ltsima{$\; \buildrel < \over \sim \;$}
\def\simlt{\lower.5ex\hbox{\ltsima}} % < over ~
\def\gtsima{$\; \buildrel > \over \sim \;$}
\def\simgt{\lower.5ex\hbox{\gtsima}} % > over ~

\def\chandra{{\it Chandra }}

\begin{document}

\title{The X-Ray Jet of PKS~0637--752: Inverse Compton Radiation from the
Cosmic Microwave Background ?} 

\author{Fabrizio Tavecchio, Laura Maraschi}
\affil{Osservatorio Astronomico di Brera, Via Brera 28, 20121, Milano, Italy}

\author{Rita M. Sambruna} 
\affil{The Pennsylvania State University, Department of Astronomy and
Astrophysics, 525 Davey Lab, State College, PA 16802}

\author{C. Megan Urry} 
\affil{Space Telescope Science Institute, 3700 San Martin Drive,
Baltimore, MD 21218}

%\cl{TITLE AND ORDER OF AUTHORS ARE PRELIMINARY} 

\begin{abstract}

We propose that the X-ray emission detected by \chandra
from the 100-kiloparsec-scale jet of PKS~0637--752 is produced through inverse
Compton scattering of the Cosmic Microwave Background (CMB). 
We analyze the physical state of the jet and show that 
inverse Compton scattering from the CMB is consistent with 
equipartition for a moderate beaming of the emission,
with Doppler factor $\delta\sim10$. 
The power transported by the jet is then similar to that of other
powerful blazars, $L_j\sim 10^{48}$~erg~s$^{-1}$, and the jet has 
low radiative efficiency. 
The radiative cooling times of the electrons are a few thousand years,
compatible with the size of the knot. 
The low-energy cutoff of the electron distribution is constrained to
be $\gamma_{\rm min} \sim 10$, the first such constraint from
spectral considerations.
A parallel analysis for the
SSC model yields far less reasonable physical conditions.

\end{abstract} 

\section{Introduction} 

The discovery made by \chandra of a handful of X-ray emitting 
kiloparsec-scale jets opens a new window on the study of the 
physics of relativistic jets. The X-ray emission offers the possibility
to observe directly the sites of particle acceleration, and allows us to
calculate important physical quantities such as the transported power. 
For this purpose it is of primary importance to establish the 
mechanism responsible for the observed X-ray emission.

Recently Chartas et al. (2000) and Schwartz et al. (2000) reported the
discovery of the X-ray emission from the jet of a distant powerful
quasar PKS~0637--752 ($z=0.651$).
From the shape of the spectral energy distribution,
one can exclude that the origin of the X-rays
is synchrotron emission from the same electron
population responsible for the radio continuum. This is because the
observed optical flux is well below (about a factor of 10) the line
connecting smoothly the radio and X-ray emission.
Schwartz et al. (2000), in discussing several possibilities for producing
the observed X-rays from knot WK7.8 (notation from Schwartz et al. 2000),
exclude bremsstrahlung emission and inverse-Compton scattering of
photons originating in the quasar. The synchrotron self-Compton (SSC)
scenario implies a magnetic field well below the equipartition value,
so Schwartz et al. conclude that a viable
explanation is de-beamed SSC emission. 
However, this solution also presents problems which we address here. 

In this letter we discuss the possibility that X-rays are
produced through inverse Compton scattering of photons of the 
Cosmic Microwave Background (CMB). The observed superluminal motion
in the jet on parsec scales, together with the close alignment
between parsec and kiloparsec scale jets 
(e.g., Tingay et al. 1998, Schwartz et al. 2000) 
suggests that the kiloparsec scale emission could be 
beamed, with a Doppler factor as high as 10-15.
This effect was possibly not taken into account in the CMB model 
discussed by Chartas et al. (2000; see their Fig.~8). 
If the X-ray jet of PKS~0637--752 is indeed sufficiently Doppler beamed,
inverse Compton scattering of CMB photons 
can easily explain the observed X-ray emission. 
In \S~2 we discuss  in detail  the implications of an EC/CMB vs an SSC model and in
\S~3 we summarize our conclusions. Througout the Letter we assume
$H_\circ =50$ km s$^{-1}$ Mpc$^{-1}$ and $q_\circ =0$.

%\section{Determining the X-Ray Emission Mechanism} 

\section{Synchrotron Self-Compton versus the External Compton CMB Model}

\subsection {General Constraints}

Given the observed spectral energy distribution of knot WK7.8 in PKS~0637--752,
the most plausible mechanism responsible for the X-ray emission is 
inverse Compton scattering, although other mechanisms, 
such as synchrotron emission
from a second electron population can not be ruled out. 
Possible sources of soft photons include the synchrotron photons 
themselves (SSC) or a radiation field external to the jet (External 
Compton, or EC). 
Given the large distance of the jet from the nucleus,
we can exclude that an important contribution comes from the 
stellar and nuclear radiation of the quasar (Schwartz et al. 2000).

Among other possible external sources of soft photons,
the CMB is particularly important. Its energy density, 
at the redshift of the source, is a factor $(1+z)^4 \simeq 7.5$
times the local value. More importantly, 
in the frame of the relativistic electrons in the jet,
it is enhanced by a factor $\Gamma ^2$,
where $\Gamma $ is the bulk Lorentz factor of the emitting plasma
(e.g., Ghisellini et al. 1998). 

For any SSC or EC model, it is possible to constrain the allowed values 
of magnetic field and Doppler beaming factor 
using standard formulae (e.g., Tavecchio, Maraschi \& Ghisellini 1998). 
We assume that the radiating region is
approximately spherical, with radius $R\sim 10^{22}$~cm 
(corresponding to $\sim$0.3~arcsec; Schwartz et al. 2000), 
filled with tangled magnetic field with intensity $B$, 
and characterized by Doppler factor 
$\delta \equiv (\Gamma [1 - \beta \cos \theta])^{-1}$.
The particle distribution is a power law with index $n=2\alpha +1$
(i.e., $N(\gamma) = K \gamma^{-n}$), 
where $\alpha =0.8$ is the observed radio spectral slope (Schwartz et
al. 2000; $F_\nu \propto \nu^{-\alpha}$), and with low and high energy
cut-off $\gamma_{\rm min}$ and $\gamma_{\rm max}$, respectively. 
Note that the X-ray spectrum also has this spectral index, $\alpha\sim0.8$
(Chartas et al. 2000), as expected in either EC or SSC models.

Using standard expressions for synchrotron and inverse Compton emissivities, 
corrected for Doppler beaming,
and using the observed radio, optical, and X-ray fluxes as constraints,
for both models the allowed values of $B$ and $\delta$ describe a line.
These are shown in Figure~1 for the SSC and EC/CMB models. 
Also shown in Figure~1 are 
lines corresponding to equipartition between 
magnetic field and radiating particles for two values of the assumed
$\gamma_{\rm min}$ (the line is rather insensitive to the value of
$\gamma _{\rm max}$). 

The SSC line depends primarily on the ratio of observed X-ray to radio 
luminosity. 
The condition for the EC/CMB model has a different slope due to the role
of additional beaming of the CMB radiation in
the rest frame of the emitting plasma --- i.e., the electrons ``see'' a beamed CMB flux
from the forward direction (e.g., Dermer 1995, Ghisellini et al. 1998).
%This beaming plays a very important role. 
%It is responsible for the completely different slopes
%of the SSC and EC/CMB lines in the $B$-$\delta$ plane.
For the SSC emission, a greater degree of beaming implies
a lower synchrotron energy density in the jet frame, which in turn
implies lower values of the magnetic field for a given Compton/synchrotron
luminosity ratio (solid line in Fig.~1).
In the EC/CMB model (dashed line in Fig.~1), in contrast,
the amplification of the external radiation increases with $\delta$,
implying a larger energy density of soft photons in the jet frame, and thus a
higher magnetic field for a given $L_C/L_S$.

From Figure~1 it is clear that the SSC model implies a very large 
deviation from equipartition unless the source is significantly 
de-beamed, with $\delta \leq 0.1$
(Schwartz et al. 2000), while the EC/CMB model is
compatible with equipartition for a moderate value of the Doppler factor,
$\delta \sim 10$.
This is consistent with measurements of superluminal motion in the
PKS~0637--752 jet on VLBI scales (Lovell 2000), 
$\Gamma > 18$ and $\theta <6.4^\circ$ (implying $\delta > 8$).

\subsection{Reproducing the Spectral Energy Distribution}

We computed spectral energy distributions (SED) for both
the synchrotron plus EC/CMB and the synchrotron plus SSC models
for a homogeneous sphere filled with
relativistic electrons (as described by Tavecchio et al. 1998). 
The CMB spectrum is modeled as a blackbody and the Compton scattering
calculation takes into account the full Klein-Nishina cross-section.
In Figure~2 we compare our best results with the 
observed radio, optical and X-ray fluxes
from knot WK7.8 (from Chartas et al. 2000).

The EC/CMB spectrum is obtained using 
equipartition values for the parameters,
$B=1.5\times10^{-5}$~G and $\delta=10$ (see Fig.~1).
For the small angles implied by VLBI observations, $\theta\lesssim6^\circ$,
we can also assume $\delta\sim\Gamma$.
The electron distribution is described by a power law with 
low- and high-energy cutoffs, $\gamma_{\rm min}=10$ 
and $\gamma_{\rm max}=4\times 10^5$, 
and normalization $K=6\times 10^{-5}$~cm$^{-3}$.
The value of $\gamma_{\rm min}$ was chosen so that the low energy
spectral break of the EC/CMB component lies in the range 
$10^{16} {\rm Hz}\lesssim \nu_{\rm break} \lesssim 10^{17} {\rm Hz}$.
We could fit the spectrum with
values in the range $5 \lesssim \gamma_{\rm min} \lesssim 20$;
for lower $\gamma_{\rm min}$ the EC emission would over-predict 
the observed optical flux, while for higher values
the X-ray spectrum would have a very steep slope.
The value of $\gamma_{\rm max}$ is fixed by the optical point and by the
values of $B$ and $\delta$. 
For the above parameters the SSC contribution is several orders of
magnitude below the EC/CMB emission and is not visible in the
figure. 
The predicted X-ray spectral slope matches well the observed slope
($\alpha_{\rm X}=0.8$; Chartas et al. 2000).

We calculated the SSC spectrum for the marginally unbeamed case,
$\delta=1$ and $B=1.2\times10^{-6}$~G (see Fig.~1), 
as a sort of ``best case'' scenario. 
For larger beaming ($\delta>1$), the magnetic field is even farther from 
equipartition ($B<10^{-6}$~G), while for lower Doppler factor (a very
de-beamed jet), the jet may approach equipartition but the total 
implied energy grows very large (see further discussion below).
The low- and high-energy cut-offs of the electron distribution are 
$\gamma_{\rm min}=2.5\times 10^3$ and $\gamma_{\rm max}=4\times 10^6$,
respectively, and the normalization is $K=19$~cm$^{-3}$.
The high value of $\gamma_{\rm min}$ is imposed by the necessity
of truncating the SSC spectrum below the soft X-ray band 
in order not to overproduce the optical flux. If, as in Fig 2, the
 low-energy tail of the
SSC spectrum gives a significant contribution to the optical flux
a strong UV emission from the jet would be expected. Note also that the
predicted X-ray spectrum is somewhat harder than actually observed.

\section {Discussion: The Power of the Jet}

Given $B$ and $\delta$, together with the observed synchrotron flux,
we can calculate the total kinetic power transported by the jet: 
\begin{equation}
L_j=\pi R^2 \Gamma^2 \beta c (U_p+U_e+U_B) ~,
\end{equation}
\noindent
(e.g., Celotti, Padovani \& Ghisellini 1997), where $U_p$, $U_e$, and $U_B$
are the energy densities of protons, electrons, and magnetic field,
respectively.

The constraints are quite interesting in the case of the EC/CMB model, 
because most of the jet energy is in the less relativistic electrons
(given their steep energy distribution), and it is those electrons
that produce most of the X-ray emission. 
For minimum electron Lorentz factor $\gamma_{\rm min}=10$ 
as derived above,
and assuming equal numbers of electrons and (cold) protons, we find 
$L_j= 3 \times 10^{48}$~erg~s$^{-1}$, 
consistent with the jet power estimated for the most powerful blazars 
(e.g., Celotti, Padovani \& Ghisellini 1997; Tavecchio et al 2000).
This is also consistent with the kinetic luminosity required to
power the lobes of giant radio galaxies (Rawlings \& Saunders 1991).
%Scarpa \& Urry 2000).

In contrast, the SSC model requires an unreasonably high value of 
the jet power. % another argument against this intepretation.
For $\delta=1$, 
%(and hence $B\simeq 10^{-6}$~G, see Fig.~1),
even if $\Gamma=2$ (which would imply $\theta\sim 60^\circ$) 
and for the lowest possible electron cutoff energy,
$\gamma_{\rm min}=2\times 10^3$ from the spectral constraints
discussed above, we find $L_j > 10^{49}$~erg~s$^{-1}$ irrespective of the
proton contribution since the average energy of the electrons is larger
than the proton mass. For the de-beamed case the situation
becomes far worse. Also, the de-beamed model implies that the jet, the
projection of which is very well aligned from VLBI to kiloparsec scales,
must actually bend away from us through a 90-degree angle. 

In the EC/CMB model the radiative power of the knot is
dominated by the $\gamma$-ray peak. Its beaming-corrected value (Sikora
et al. 1997) is $P_{\rm rad} \simeq 6 \times 10^{43}$~erg~s$^{-1}$. The
position of the $\gamma$-ray peak, $\nu_{\rm peak} \simeq \nu_{\rm CMB}
\delta^2 \gamma_{\rm max}^2$, is very well constrained because the CMB peak
frequency ($\nu_{\rm CMB}$) is known and the values of the Doppler
factor ($\delta$) and the maximum Lorentz factor ($\gamma_{\rm max}$)
are well constrained by the assumption of equipartition. 
The corresponding $\gamma$-ray power for the minimally beamed SSC model
($\delta=1$) is $P_{\rm rad} = 6.7\times 10^{45}$~ergs~s$^{-1}$, and this
value increases rapidly with decreasing $\delta$. 

Even for the EC/CMB model, the kinetic power of the jet is very high
compared to the observed luminosity, implying very low radiative efficiency. 
(The efficiency is higher for the SSC model and it increases as
$\delta $ decreases.)
The electrons emitting via synchrotron in the optical band and 
via EC/CMB in the $\gamma$-rays have relatively short lifetimes, 
$t_{\rm cool} \sim 10^{11}$~s (Compton cooling dominates 
synchrotron cooling) and can travel for at most
$d \lesssim \Gamma c t_{\rm cool} \sim 10$~kpc.
This is consistent with ($\sim 3$ times) the size of the knot, 
which can therefore arise from a single acceleration site.
%Note that for the SSC model, the electron lifetimes are far shorter,
%$t_{\rm cool} \sim 10^{XX}$~s,
%CHECK!
%so reacceleration within the knot is probably required.
Given the close alignment of the jet with the line of sight, 
its minimum de-projected length is $\sim1$~Mpc, and
{\it in situ} reacceleration beyond knot WK7.8 is required in either case.

We conclude, from spectral, equipartition and jet power arguments, that 
inverse Compton scattering of CMB photons is the most likely emission 
mechanism for the observed X-rays from PKS~0637--752.

\section{Conclusion}

We have shown that the X-ray emission from knot WK7.8 in the 
PKS~0637--752 jet likely originates from upscattering of CMB photons 
provided the jet is still relativistic on kiloparsec scales.
The physical parameters in the emitting region are consistent 
with equipartition for moderate values of the Doppler
beaming factor, $\delta \sim 10$, which also agrees with 
observed superluminal motion in this jet (albeit on much smaller scales).
Given a high degree of relativistic beaming, the size of the knot 
is also consistent with the diffusion length for the radiating electrons.
%For the SSC model, in contrast, the electron lifetimes are much shorter,
%and reacceleration within the knot would probably be required.

Modeling the radio-to-X-ray continuum constrains the value of 
the minimum Lorentz factor of the emitting electrons to be
$\gamma_{\rm min} \sim 10$. 
This is the first such constraint from spectral modeling. Generally,
the synchrotron emission from these electrons is self absorbed
(here, $\nu_{\rm abs} \sim 10^6$~Hz). 
Minimum values for electron energy have been inferred from general
arguments (e.g., Reynolds et al. 1996) or
the lack of Faraday depolarization (e.g., Wardle et al. 1998), but a
direct probe of $\gamma_{\rm min}$ is possible only 
when the observed spectrum describes a sharp break in the
Compton-scattered radiation. 

The estimated kinetic power of the jet, $L_j \sim 10^{48}$~ergs~s$^{-1}$,
is then consistent with that of other powerful blazar jets. 
The SSC model requires far less reasonable assumptions, notably
that the parsec-scale jet is nearly aligned with the line of sight but
the kiloparsec-scale jet is roughly in the plane of the sky,
and implies a jet kinetic power at least one order of magnitude larger.
Observations of the jet in the UV could help discriminating between the two
possibilities.

With the favored EC/CMB model (and also in the SSC case), 
the X-ray jet has remarkably low radiative efficiency, 
suggesting PKS~0637--752 should have powerful extended radio lobes or be in an early
phase of evolution.

\acknowledgements
FT is grateful to STScI for hospitality during the 
preparation of this work. This work was supported in part by
NASA grant NAG5-9327. FT and LM acknowledge partial support from grants
CEE ERBFMRX-CT98-0195 and MURST-COFIN-98-02-15-41. RMS acknowledges
support from NASA contract NAS--38252.

\clearpage

\vskip 1.5 truecm

\centerline{ \bf Figure Captions}

\figcaption[lc]{Allowed values of magnetic field and Doppler beaming factor
for knot WK7.8 of PKS~0637--752. 
({\it Solid line:}) allowed values of $B$ and $\delta$ if
the observed X-ray emission is produced by a synchrotron self-Compton model.
({\it Dashed line:}) allowed values if the X-rays come from inverse Compton
scattering of cosmic microwave background photons (the EC/CMB model).
({\it Dotted lines:}) allowed values of $B$ and $\delta$ under the assumption of
equipartition between radiating particles and magnetic field for two
different values of the low energy cut-off of the electron distribution,
$\gamma_{\rm min}=10$ and $\gamma_{\rm min}=10^3$. 
The EC/CMB emission can clearly be consistent with equipartition 
conditions for moderately large values of the Doppler factor 
($\delta \sim 10$), while the SSC case is far from equipartition 
unless an enormous degree of de-beaming is present
($\delta \ll 0.1$).}

\figcaption[lc]{Spectrum of knot WK7.8 of the PKS~0637--752 jet.
({\it Filled circles:}) Observed fluxes taken from Chartas et al. 
(2000) and references therein. 
({\it Solid line:}) The EC/CMB model, in which cosmic microwave background
photons are inverse Compton scattered by the synchrotron-emitting electrons.
The emission region
is assumed to be approximately spherical with radius $R=10^{22}$~cm,
the magnetic field intensity is $B=1.5\times10^{-5}$~G and we assume a
Doppler factor $\delta \sim \Gamma =10$ (consistent with an observing
angle $\theta \sim 6^\circ$). The electron
distribution has a power-law shape with extremes 
$\gamma_{\rm min}=10$ and $\gamma_{\rm max}=4\times 10^5$, 
normalization $K=6\times 10^{-5}$~cm$^{-3}$,
and slope 2.6 ($\alpha = 0.8$). 
The external radiation is a blackbody with the CMB temperature and 
energy density.
({\it Dashed line:}) SSC model. The emission region is a sphere with the
same radius as in the EC/CMB model, the magnetic field intensity is
$B=1.25\times10^{-6}$~G and we assume a Doppler factor $\delta=1$. The
electron distribution has a power-law shape with extremes 
$\gamma_{\rm min}=2.5\times 10^3$ and $\gamma_{\rm max}=4\times 10^6$, 
normalization $K=19$~cm$^{-3}$, and slope 2.6. Note the break in the
synchrotron spectrum at $\sim 100$ MHz due to the high value of
$\gamma_{\rm min}$.}

\newpage

\begin{figure}
%\centerline{\plotone{xsed_mio.ps}}
\centerline{\plotone{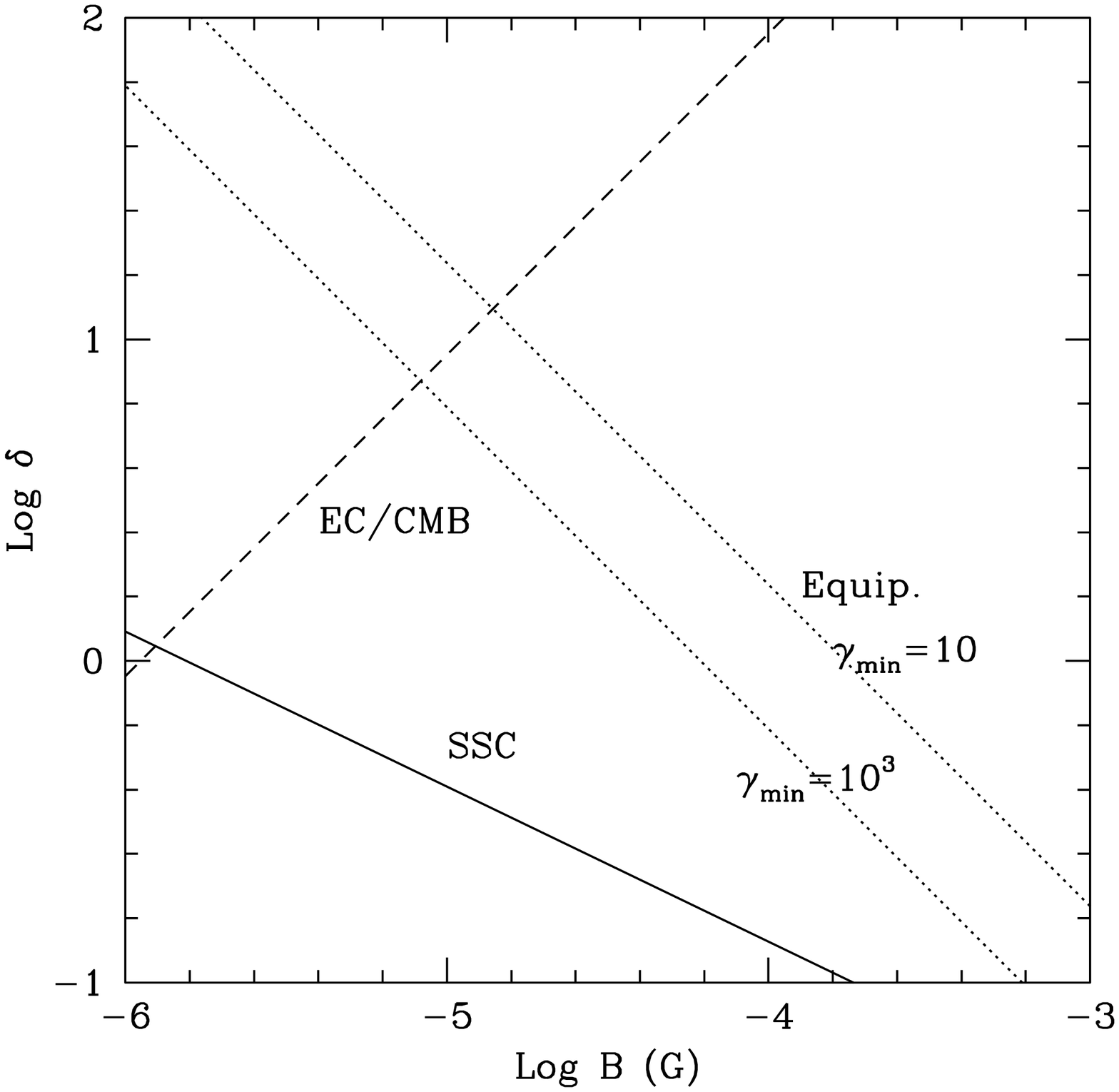}}
\end{figure}

\begin{figure}
%\centerline{\plotone{xsed_mio.ps}}
\centerline{\plotone{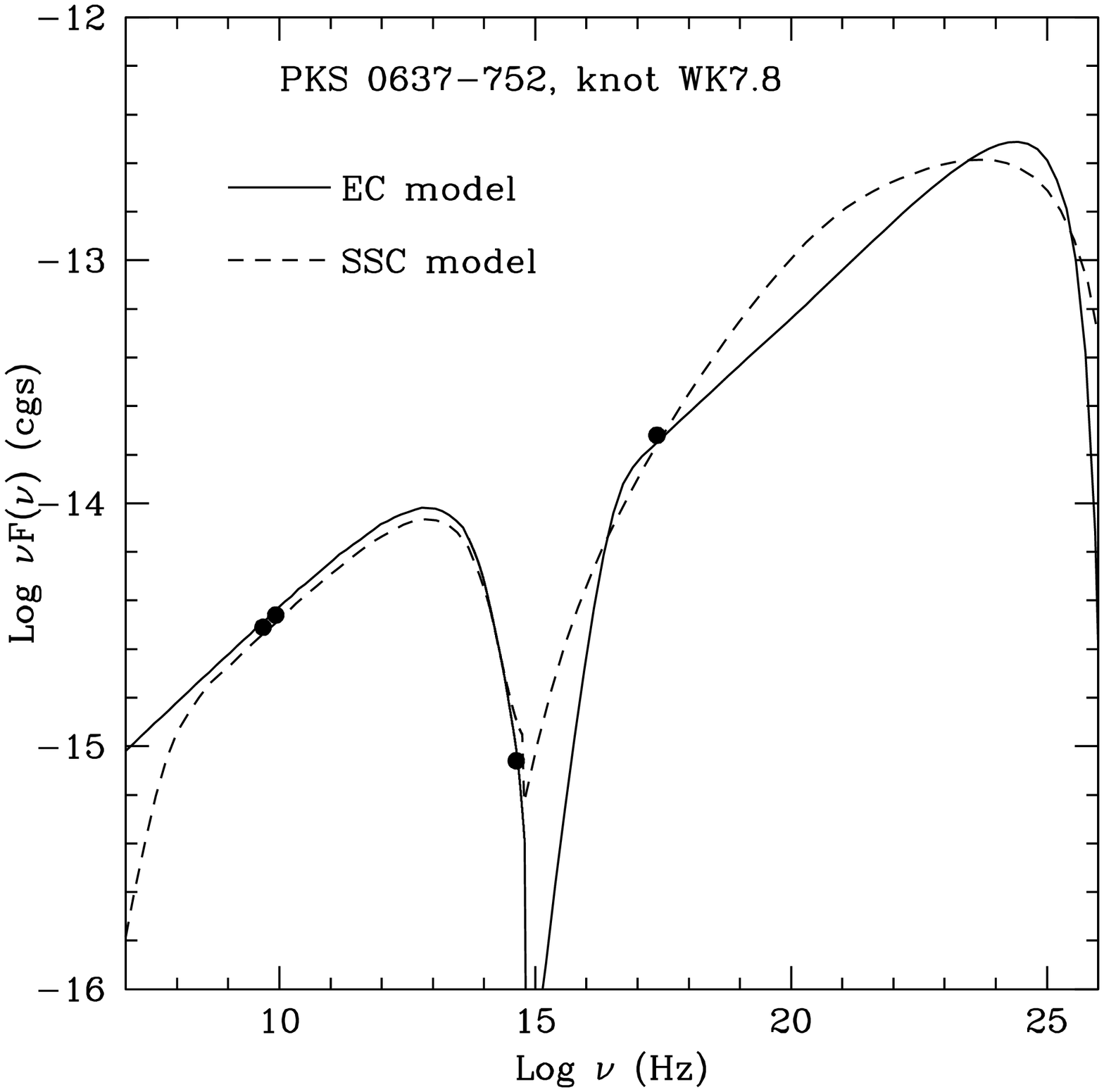}}
\end{figure}


\newpage 

\begin{references}

\reference {} Celotti, A., Padovani, P., \& Ghisellini, G. 1997, MNRAS,
286, 415

\reference {} Chartas, G., et al. 2000, ApJ, in press (astro-ph/0005227) 

\reference {} Dermer, C.D., 1995, ApJ, 446, L63

\reference {} Ghisellini, G., et al. 1998, MNRAS, 301, 451

\reference {} Lovell, J.E., 2000, in Astrophysical Phenomena Revealed by
Space VLBI, eds. H. Hirabayashi, P.G. Edwards and D.W. Murphy, 215

\reference {} Rawling, S. \& Saunders, R., 1991, Nature, 349, 138

\reference {} Reynolds, C.S., et al. 1996, MNRAS, 283, 873

\reference {} Schwartz, D.A., et al. 2000, submitted to ApJ (astro-ph/0005255)

\reference {} Sikora, M., et al. 1997, ApJ, 484, 108

\reference {} Tavecchio, F., et al. 2000, ApJ, in press (astro-ph/0006443)

\reference {} Tingay, S.J., et al. 1998, ApJ, 497, 594

\reference {} Wardle, J. F. C., Homan, D. C., Ojha, R., Roberts,
D. H. 1998, Nature, 395, 457

\end{references}
\end{document}